\documentclass[twocolumn]{aastex61}

\usepackage{amsmath}
\usepackage{amsfonts}
\usepackage{graphicx}
\usepackage{natbib}
\usepackage{txfonts}
\usepackage{color}
\usepackage{wrapfig}
\usepackage{amssymb}
\usepackage{color}

\shorttitle{A multiline study of a high-mass YSO in the SMC} 
\shortauthors{T. Shimonishi et al.} 

\begin{document}

\title{A multiline study of a high-mass young stellar object in the Small Magellanic Cloud with ALMA: The detection of methanol gas at 0.2 solar metallicity}

\correspondingauthor{Takashi Shimonishi} 
\email{shimonishi@astr.tohoku.ac.jp} 

\author{Takashi Shimonishi} 
\affiliation{Frontier Research Institute for Interdisciplinary Sciences, Tohoku University, Aramakiazaaoba 6-3, Aoba-ku, Sendai, Miyagi, 980-8578, Japan} 
\affiliation{Astronomical Institute, Tohoku University, Aramakiazaaoba 6-3, Aoba-ku, Sendai, Miyagi, 980-8578, Japan} 

\author{Yoshimasa Watanabe} 
\affiliation{Faculty of Pure and Applied Sciences, University of Tsukuba, 1-1-1, Tennodai, Tsukuba, Ibaraki 305-8577, Japan}
\affiliation{Tomonaga Center for the History of the Universe, University of Tsukuba, Tsukuba, Ibaraki 305-8571, Japan} 

\author{Yuri Nishimura} 
\affiliation{Institute of Astronomy, The University of Tokyo, 2-21-1, Osawa, Mitaka, Tokyo 181-0015, Japan}
\affiliation{National Astronomical Observatory of Japan, 2-21-1, Osawa, Mitaka, Tokyo 181-8588, Japan}

\author{Yuri Aikawa} 
\affiliation{Department of Astronomy, Graduate School of Science, The University of Tokyo, 7-3-1 Hongo, Bunkyo-ku, Tokyo 113-0033, Japan}

\author{Satoshi Yamamoto} 
\affiliation{Department of Physics, The University of Tokyo, 7-3-1, Hongo, Bunkyo-ku, Tokyo 113-0033, Japan}
\affiliation{Research Center for the Early Universe, The University of Tokyo, 7-3-1, Hongo, Bunkyo-ku, Tokyo, 113-0033, Japan}

\author{Takashi Onaka} 
\affiliation{Department of Astronomy, Graduate School of Science, The University of Tokyo, 7-3-1 Hongo, Bunkyo-ku, Tokyo 113-0033, Japan}

\author{Nami Sakai} 
\affiliation{RIKEN, 2-1 Hirosawa, Wako, Saitama 351-0198, Japan}

\author{Akiko Kawamura} 
\affiliation{National Astronomical Observatory of Japan, 2-21-1, Osawa, Mitaka, Tokyo 181-8588, Japan}



\begin{abstract} 
We report the results of subparsec-scale submillimeter observations towards an embedded high-mass young stellar object in the Small Magellanic Cloud (SMC) with ALMA. 
Complementary infrared data obtained with the \textit{AKARI} satellite and the Gemini South telescope are also presented. 
The target infrared point source is spatially resolved into two dense molecular cloud cores; one is associated with a high-mass young stellar object (YSO core), while another is not associated with an infrared source (East core). 
The two cores are dynamically associated but show different chemical characteristics. 
Emission lines of CS, C$^{33}$S, H$_2$CS, SO, SO$_2$, CH$_3$OH, H$^{13}$CO$^{+}$, H$^{13}$CN, SiO, and dust continuum are detected from the observed region. 
Tentative detection of HDS is also reported. 
The first detection of CH$_3$OH in the SMC has a strong impact on our understanding of the formation of complex organic molecules in metal-poor environments. 
The gas temperature is estimated to be $\sim$10 K based on the rotation analysis of CH$_3$OH lines. 
The fractional abundance of CH$_3$OH gas in the East core is estimated to be (0.5--1.5) $\times$ 10$^{-8}$, which is comparable with or marginally higher than those of similar cold sources in our Galaxy despite a factor of five lower metallicity in the SMC. 
This work provides observational evidence that an organic molecule like CH$_3$OH, which is largely formed on grain surfaces, can be produced even in a significantly lower metallicity environment compared to the solar neighborhood. 
A possible origin of cold CH$_3$OH gas in the observed dense core is discussed. 
\end{abstract} 




\section{Introduction} \label{sec_intro} 
%Astrochemistry at low metallicity 
Interstellar chemistry at low metallicity is crucial to understand chemical processes in past galaxies since cosmic metallicity is increasing in time with the evolution of our universe. 
Observational studies of dense molecular gas in nearby low-metallicity galaxies play an important role in understanding chemistry in star-forming regions as a function of galactic metallicity. 
Chemical compositions of molecular cloud cores and protostellar envelopes are particularly interesting for understanding initial chemical conditions of star-formation in metal-poor environments. 

%Previous studies on the LMC
Characteristic chemical compositions around high-mass young stellar objects (YSOs) in low metallicity environments are emerging thanks to recent advances in infrared and radio observations towards compact sources in the Large Magellanic Cloud. 
The abundance of organic molecules such as CH$_3$OH is one of the key issues in recent astrochemical studies toward low metallicity systems. 
Infrared observations of ices suggest that solid CH$_3$OH around embedded high-mass YSOs in the LMC is less abundant than those of Galactic similar objects \citep{ST16}. 
In single-dish radio observations, CH$_3$OH lines are detected from two LMC star-forming regions \citep{Hei99,Wan09}. 
Spectral line surveys and abundance analyses towards seven molecular clouds in the LMC suggest that CH$_3$OH gas is less abundant in the LMC \citep{Nis16}. 
Searches for maser emission in star-forming regions reported the lower detection rate of of CH$_3$OH masers in the LMC compared to those in our Galaxy \citep[e.g.,][]{Gre08,Ell10}. 
Submillimeter observations with the Atacama Large Millimeter/submillimeter Array (ALMA) reported that gas-phase CH$_3$OH towards a hot molecular core in the LMC is underabundant by 1--3 orders of magnitude compared to those towards Galactic hot cores \citep{ST16b}. 
On the other hand, \citet{Sew18} recently detected CH$_3$OH and other complex organic molecules toward two different hot molecular cores in the LMC. 
They argue that the abundances of these molecules are scaled by the metallicity of the LMC and comparable to those found at the lower end of the range in Galactic hot cores.
Chemical processes of organic molecules in low metallicity environments are still an open question and observational tests for further metal-poor samples are essentially important.

%SMC
The Small Magellanic Cloud (SMC), the prime target of this paper, is one of the nearest star-forming galaxies to our Galaxy \citep[62.1 $\pm$ 2.0 kpc,][]{Gra14}. 
It is well known that the SMC has a low-metallicity environment, which is about one-fifth of the solar neighborhood \citep[e.g.,][]{Duf84,Rus92,Ven99}. 
Relative abundances among heavy elements are also different in the SMC, e.g., the ratio of elemental nitrogen relative to carbon or oxygen is lower in the SMC by a factor of $\sim$3 \citep{Rus92}. 
A dust-to-gas ratio is also reported to be lower in the SMC than in our Galaxy by a factor of 5 or more \citep[e.g.,][]{Ler07,Rom17}. 
A lower abundance of dust grains leads to harsh radiation environment, which results in elevated dust/gas temperatures and more effective photoprocessing of dense interstellar medium \citep[e.g.,][]{Agu03,vanL10_b}. 
The above environmental differences would affect chemical compositions in molecular clouds in the SMC as in the case of the LMC. 

%Previous molecular line studies towards the SMC 
Pioneering works on molecular gas multiline studies for the SMC are presented in \citet{Chi97,Chin98} and \citet{Hei99} based on observations with the 15-m Swedish-ESO Submillimetre Telescope. 
These studies reported detection of various molecular species, including CO, CS, SO, C$_2$H, HCO$^+$, HCN, HNC, H$_2$CO, and C$_3$H$_2$ towards active star-forming regions N27 and LIRS 36. 
Recently, \citet{Mur17} reported high-spatial resolution observations of CO, $^{13}$CO, C$^{18}$O, and CS with ALMA towards a star-forming region, N83C. 
They revealed subparsec-scale structures and physical properties of molecular clouds in the SMC. 
High spatial resolution molecular line observations towards the SMC are, however, still limited to a few molecular species, and the chemistry of star-forming cores in the SMC remains to be investigated. 

%Ices in the SMC 
Infrared observations of ices toward embedded high-mass YSOs in the SMC are also reported in the literature \citep{vanL08,Oli11,Oli13}. 
Currently, H$_2$O and CO$_2$ ices are detected in the SMC, while CO ice is not detected. 
Astrochemical simulations dedicated to the SMC environment suggest that the observed absence of CO ice in the SMC is explained by a combined effect of increased dust temperatures and reduced metal abundances \citep{Ach16,Pau18}. 

%This paper
In this paper, we report the results of subparsec-resolution submillimeter observations towards a high-mass YSO in the SMC with ALMA. 
Section \ref{sec_tarobsred} describes the details of observations and data reduction. 
The obtained spectra and images are presented in Section \ref{sec_res}. 
Analysis of the spectral data and the derivation of the physical quantities of molecular gas are also described in this section. 
Section \ref{sec_disc} describes physical and chemical properties of the observed source. 
A particular focus is given to the properties of CH$_3$OH gas in the metal-poor environment of the SMC. 
The conclusions of this paper are presented in Section \ref{sec_sum}.

\section{Observations and data reduction} \label{sec_tarobsred} 
\subsection{Target} \label{sec_tar}
The target is the infrared source IRAS 01042-7215 located in the SMC. 
It is spectroscopically identified to be an embedded high-mass YSO in previous infrared studies \citep{vanL08,Oli11,Oli13,thesis,Ruf15}. 
The luminosity of the source is estimated to be 2 $\times$ 10$^4$ L$_{\sun}$. 
More detailed discussion about the YSO's properties is given in Section \ref{sec_disc}.

\subsection{Observations} \label{sec_obs}
Observations were carried out with ALMA in November 2016 as a part of the Cycle 4 program 2016.1.00394.S (PI T. Shimonishi). 
The telescopes were pointed to RA = 01$^\mathrm{h}$05$^\mathrm{m}$49$\fs$32 and Dec = -71$^\circ$59$\arcmin$48$\farcs$4 (J2000), which corresponds to the target's infrared peak position. 
The target was observed in Band 6 with four spectral windows covering 241.48--243.36, 243.84-245.71, 256.99--258.87, and 258.85--260.72 GHz in the rest-frame frequency. 
The frequency resolution is 977 kHz before channel averaging. 
The total on-source integration time is 39 minutes. 
We use J2357-5311, J2056-4714, and J0102-7546 as a flux, bandpass, and gain calibrator, respectively. 
In total 43 antennas are used for the observations with the minimum and maximum baseline length of 19 m and 1.1 km.

\subsection{Data reduction} \label{sec_red}
Raw interferometric data are processed using the \textit{Common Astronomy Software Applications} (CASA) package. 
The calibration is done with CASA 4.7.0 and the imaging is with CASA 4.7.2. 
The synthesized beam size is 0.35--0.37$\arcsec$ $\times$ 0.22--0.23$\arcsec$ with a position angle of -25 degree, which corresponds to 0.11 pc $\times$ 0.07 pc at the distance of the SMC (62.1 kpc). 
The primary beam has a full-width at half-maximum (FWHM) of about 25$\arcsec$, and the maximum recoverable angular scale is about 3$\arcsec$. 
The primary beam correction is done by using the impbcor task in CASA. 

The spectra and continuum flux are extracted from the 0.34$\arcsec$ (0.10 pc) diameter region centered at the three positions (Position 1, 2, 3, hereafter P1, P2, P3), where the dust continuum and/or molecular emission lines have a local peak (see Section \ref{sec_img}). 
Their coordinates are listed in Table \ref{tab_pos}. 
The continuum emission is subtracted from the spectral data using the uvcontsub task in CASA. 
The velocity resolution of the extracted spectra is 1.5 km s$^{-1}$. 

%%%%%%%%%%
\begin{deluxetable*}{ l c c c c c}
\tablecaption{Positions for the spectral extraction \label{tab_pos}}
\tablewidth{0pt}
\tabletypesize{\scriptsize} 
\tablehead{
\colhead{Position}  & \colhead{R.A.}        &  \colhead{Decl.}     &  \colhead{Detected molecules}  &  \colhead{Note}  \\
\colhead{}              & \colhead{(J2000)}  &  \colhead{(J2000)} &  \colhead{}   &  \colhead{}            \\
}
\startdata 
P1       &   01 05 49.33    &    -71 59 48.46    & CS, SO, H$^{13}$CO$^{+}$                                       &  Continuum, H$^{13}$CO$^+$, and infrared  peak    \\ 
P2       &   01 05 49.54    &    -71 59 48.94    & CS, C$^{33}$S, H$_2$CS, SO, SO$_2$, (HDS),       &  C$^{33}$S, H$_2$CS, and SiO peak    \\ 
            &                         &                             & CH$_3$OH, H$^{13}$CN, H$^{13}$CO$^{+}$, SiO   &   \\
P3       &   01 05 49.50    &    -71 59 49.41    & CS, SO, SO$_2$, CH$_3$OH, H$^{13}$CO$^{+}$    &  CH$_3$OH peak    \\ 
\enddata
\tablecomments{Tabulated positions are shown in Figure \ref{images}. HDS is a tentative detection. } 
\end{deluxetable*}
%%%%%%%%%%

\subsection{Complementary infrared data} \label{sec_ir} 
Complementary infrared datasets are also presented in this work. 
One contains high spatial resolution mid-infrared imaging data obtained with T-ReCS at the Gemini South telescope (Program ID: S10B-120, PI: T. Shimonishi). 
The observations were carried out using a broad-band filter in the N-band (7.70--12.97 $\mu$m, centered at 10.36 $\mu$m) as well as six narrow-band filters (centered at 7.73, 8.74, 9,69, 10.38, 11.66, and 12.33 $\mu$m). 
The spatial resolution of the imaging data is about 0.55$\arcsec$ according to the FWHM measurement for a photometric standard star. 

Another complementary data set is a near-infrared 2--5 $\mu$m spectrum obtained with the Infrared Camera (IRC) on board \textit{AKARI} \citep{Mur07,TON07}. 
The observations were carried out using the slitless NG spectroscopy mode with the Np point source aperture (R $\sim$80) as a part of the \textit{AKARI} post-helium mission program (Program ID: LMCNG, PI: T. Onaka). 
The near-infrared spectrum is extracted from the infrared peak position. 
A typical FWHM of point sources in the \textit{AKARI} data is approximately 5$\arcsec$--8$\arcsec$.

\section{Results} \label{sec_res} 
\subsection{Spectra} \label{sec_spc} 
\begin{figure*}[tbp]
\begin{center}
\includegraphics[width=15cm]{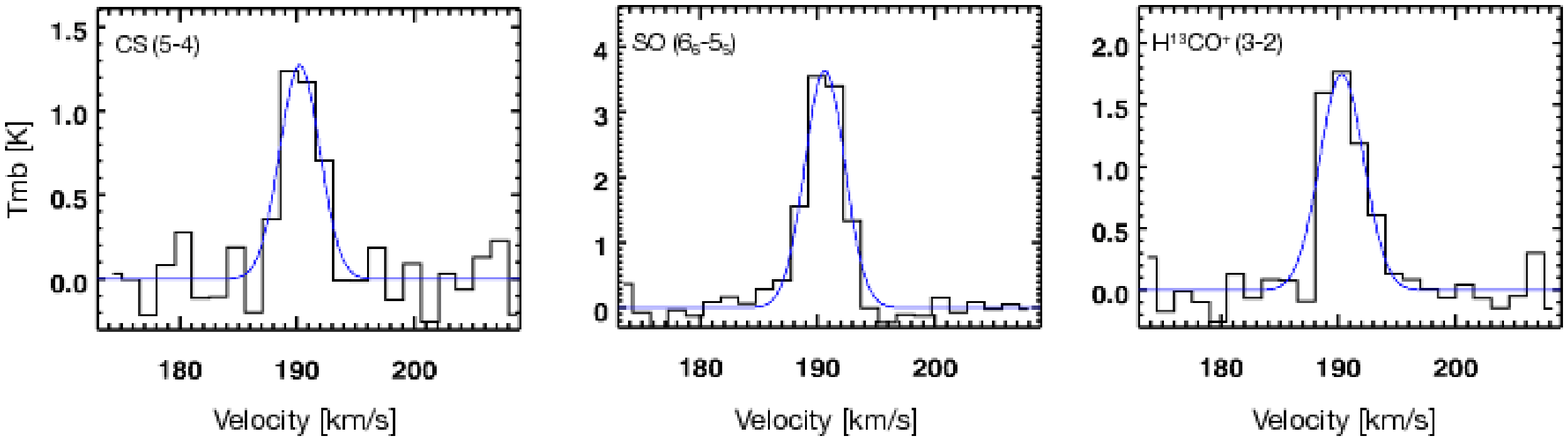}
\caption{
Spectra extracted at Position 1. 
Emission lines of CS, SO, and H$^{13}$CO$^+$ are shown. 
The blue lines represent Gaussian profiles fitted to the spectra. 
}
\label{spec_pos1}
\end{center}
\end{figure*}

\begin{figure*}[tp]
\begin{center}
\includegraphics[width=15cm]{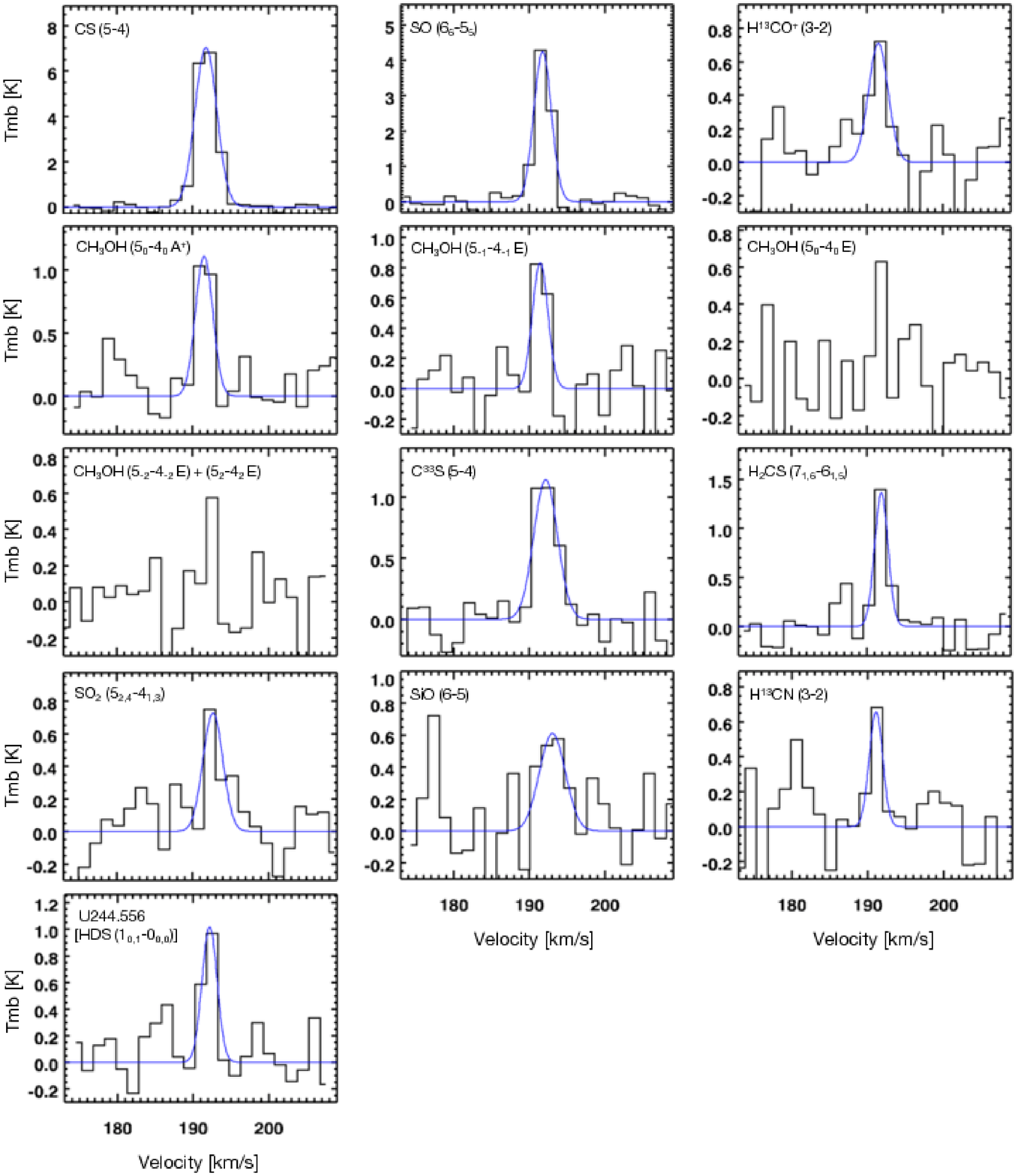}
\caption{
Spectra extracted at Position 2. 
Emission lines of CS, SO, H$^{13}$CO$^+$, CH$_3$OH, C$^{33}$S, H$_2$CS, SO$_2$, SiO, H$^{13}$CN, and U244.556 (tentatively attributed to HDS) are shown. 
The blue lines represent Gaussian profiles fitted to the spectra. 
For CH$_3$OH (5$_{0}$--4$_{0}$ E) and (5$_{\pm2}$--4$_{\pm2}$ E) lines, integrated intensities are derived by directly integrating the spectrum between 190 km s$^{-1}$ and 194 km s$^{-1}$. 
}
\label{spec_pos2}
\end{center}
\end{figure*}

\begin{figure*}[tbp]
\begin{center}
\includegraphics[width=15cm]{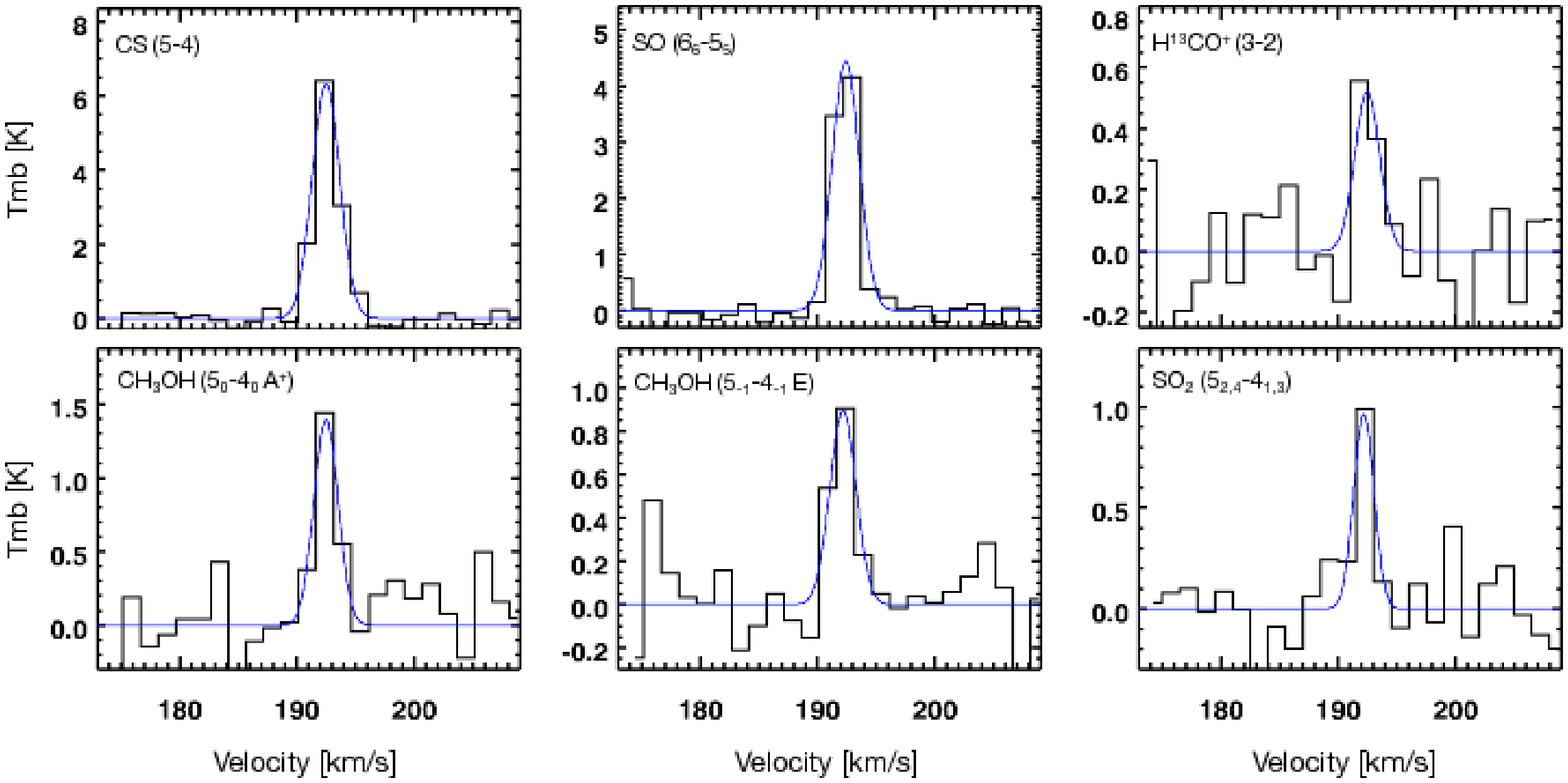}
\caption{
Spectra extracted at Position 3. 
Emission lines of CS, SO, H$^{13}$CO$^+$, CH$_3$OH, and SO$_2$ are shown. 
The blue lines represent Gaussian profiles fitted to the spectra. 
}
\label{spec_pos3}
\end{center}
\end{figure*}

Figures \ref{spec_pos1}--\ref{spec_pos3} show spectra extracted from the three positions described in Section \ref{sec_red}. 
Spectral lines are identified with the aid of the Cologne Database for Molecular Spectroscopy\footnote{https://www.astro.uni-koeln.de/cdms} \citep[CDMS,][]{Mul01,Mul05}. 
The detection criteria adopted here are the 3$\sigma$ significance level and the velocity coincidence with the systemic velocity of nearby clouds \citep[between 186 km s$^{-1}$ and 196 km s$^{-1}$,][]{Isr93}. 
Line parameters are measured by fitting a single Gaussian profile to observed lines. 
In some cases we subtract a local baseline which is estimated from adjacent line-free regions by fitting a polynomial to correct for small baseline ripples. 
Based on the fitting, we estimate a peak brightness temperature, an FWHM, an LSR velocity, and an integrated intensity for each line. 
For CH$_3$OH(5$_{0}$--4$_{0}$ E) and (5$_{\pm2}$--4$_{\pm2}$ E) lines, the integrated intensities are calculated by directly integrating the spectrum between 190 km s$^{-1}$ and 194 km s$^{-1}$ because they are too narrow to fit a Gaussian. 
These two lines are attributed to tentative detections given their low intensity. 
The results of the spectral fits are shown in Figures \ref{spec_pos1}--\ref{spec_pos3}, and the measured line parameters are summarized in Table \ref{tab_lines}. 
The table also contains the estimated upper limits on several important non-detection lines. 

Molecular emission lines of CS, C$^{33}$S, H$_2$CS, SO, SO$_2$, CH$_3$OH, H$^{13}$CO$^{+}$, H$^{13}$CN, and SiO are detected from the observed region. 
Among these species, CH$_3$OH, SO$_2$, H$_2$CS, C$^{33}$S, and H$^{13}$CN are for the first time detected in the SMC. 
Measured line widths are typically 3--4 km s$^{-1}$. 
The spectra at P1 show slightly broader line widths compared to those at P2 and P3. 
An unidentified line is also detected near 244.5556 GHz (hereafter U244.556), which is tentatively attributed to HDS as further discussed in Section \ref{sec_uline}. 
High-excitation lines (upper-state energy $>$100 K) are covered in the present frequency settings but not detected. 
A carbon-chain-related molecule, c-C$_3$H$_2$, is not detected. 

%%%%%%%%%%
%%%%%%%%%%
%
%%%%%%%%%%
%%%%%%%%%%
%
%%%%%%%%%%
%%%%%%%%%%

%%%%%%%%%%
\begin{deluxetable*}{ l c c c c c c c c c }
\tablecaption{Line Parameters \label{tab_lines}}
\tablewidth{0pt}
\tabletypesize{\scriptsize} 
\tablehead{
\colhead{Molecule}   & \colhead{Transition}                               &       \colhead{$Eu$} &       \colhead{Frequency} &        \colhead{$T_{\mathrm{b}}$} &     \colhead{$\Delta$$V$} &     \colhead{$\int T_{\mathrm{b}} dV$} &       \colhead{$V_{\mathrm{LSR}}$} &        \colhead{RMS} &       \colhead{Note} \\
\colhead{ }          & \colhead{ }                                        &        \colhead{(K)} &           \colhead{(GHz)} &             \colhead{(K)} &          \colhead{(km s$^{-1}$)} &             \colhead{(K km s$^{-1}$)} &          \colhead{(km s$^{-1}$)} &        \colhead{(K)} &           \colhead{}
}
\startdata 
\textit{Position 1} & & & & & & & & \\
 CS                  &  5--4                                    &   35       & 244.93556       &   1.27 $\pm$   0.11  &    4.0          & 5.45 $\pm$ 1.11      & 190.2           & 0.19 &    \nodata \\
 SO                  &  6$_6$--5$_5$                            &   57       & 258.25583       &   3.64 $\pm$   0.10  &    4.0          & 15.61 $\pm$ 0.97     & 190.6           & 0.18 &    \nodata \\
 H$^{13}$CO$^+$      &  3--2                                    &   25       & 260.25533       &   1.75 $\pm$   0.10  &    4.4          & 8.24 $\pm$ 1.13      & 190.3           & 0.19 &    \nodata \\
 CH$_3$OH            &  5$_{0}$--4$_{0}$ A$^+$                  &   35       & 241.79143       &             $<$0.7   &    \nodata      &             $<$1.4   &    \nodata      & 0.20 &    \nodata \\
 SO$_2$              &  5$_{2,4}$--4$_{1,3}$                    &   24       & 241.61580       &             $<$0.4   &    \nodata      &             $<$1.3   &    \nodata      & 0.20 &    \nodata \\
 C$^{33}$S           &  5--4                                    &   35       & 242.91361       &             $<$0.4   &    \nodata      &             $<$1.3   &    \nodata      & 0.20 &    \nodata \\
 $^{33}$SO          &  6$_{7}$--5$_{6}$                  &   47       & 259.28403       &             $<$0.4   &    \nodata      &             $<$1.2   &    \nodata      & 0.19 &    (1) \\
 c-C$_3$H$_2$       &  3$_{2,1}$--2$_{1,2}$        &   18       & 244.22215       &            $<$0.4   &    \nodata      &             $<$1.2   &    \nodata      & 0.19 &    \nodata \\
\tableline 
\textit{Position 2} & & & & & & & & \\
 CS                  &  5--4                                    &   35       & 244.93556       &   7.05 $\pm$   0.12  &    3.4          & 25.46 $\pm$ 0.91     & 191.7           & 0.19 &    \nodata \\
 SO                  &  6$_6$--5$_5$                            &   57       & 258.25583       &   4.25 $\pm$   0.11  &    2.7          & 12.34 $\pm$ 0.77     & 191.7           & 0.19 &    \nodata \\
 H$^{13}$CO$^+$      &  3--2                                    &   25       & 260.25533       &   0.71 $\pm$   0.12  &    3.0          & 2.30 $\pm$ 0.96      & 191.5           & 0.21 &    \nodata \\
 CH$_3$OH            &  5$_{0}$--4$_{0}$ A$^+$                  &   35       & 241.79143       &   1.10 $\pm$   0.12  &    2.7          & 3.20 $\pm$ 0.75      & 191.5           & 0.19 &    \nodata \\
 CH$_3$OH            &  5$_{-1}$--4$_{-1}$ E                    &   40       & 241.76722       &   0.83 $\pm$   0.11  &    2.4          & 2.15 $\pm$ 0.62      & 191.4           & 0.19 &    \nodata \\
 CH$_3$OH            &  5$_{0}$--4$_{0}$ E                      &   48       & 241.70022       &   0.63 $\pm$   0.38  & 1.5    & 0.98 $\pm$ 0.34      & 191.7           & 0.19 &    (2) (3) \\
 CH$_3$OH            &  5$_{1}$--4$_{1}$ A$^-$                  &   50       & 243.91583       &             $<$0.4   &    \nodata      &             $<$1.2   &    \nodata      & 0.19 &    \nodata \\
 CH$_3$OH            &  5$_{1}$--4$_{1}$ E                      &   56       & 241.87907       &             $<$0.4   &    \nodata      &             $<$1.2   &    \nodata      & 0.19 &    \nodata \\
 CH$_3$OH            &  5$_{\pm2}$--4$_{\pm2}$ E     &   61       & 241.90416       &   0.58 $\pm$   0.38  & 1.5   & 0.83 $\pm$ 0.32      & 192.4           & 0.19 &    (2) (3) (4) \\
 CH$_3$OH            &  5$_{2}$--4$_{2}$ A$^+$                  &   73       & 241.88771       &             $<$0.4   &    \nodata      &             $<$1.2   &    \nodata      & 0.19 &    \nodata \\
 SO$_2$              &  5$_{2,4}$--4$_{1,3}$                    &   24       & 241.61580       &   0.73 $\pm$   0.10  &    3.1          & 2.39 $\pm$ 1.48      & 192.7           & 0.19 &    \nodata \\
 C$^{33}$S           &  5--4                                    &   35       & 242.91361       &   1.14 $\pm$   0.12  &    3.7          & 4.50 $\pm$ 1.01      & 192.1           & 0.19 &    \nodata \\
 $^{33}$SO             &  6$_{7}$--5$_{6}$             &   47       & 259.28403       &             $<$0.4   &    \nodata      &             $<$1.3   &    \nodata      & 0.21 &    (1) \\
 H$_2$CS             &  7$_{1,6}$--6$_{1,5}$                    &   60       & 244.04851       &   1.37 $\pm$   0.11  &    2.1          & 3.05 $\pm$ 0.64      & 191.9           & 0.19 &    \nodata \\
 SiO                 &  6--5                                    &   44       & 260.51800       &   0.61 $\pm$   0.12  &    4.0          & 2.63 $\pm$ 1.12      & 193.0           & 0.21 &    \nodata \\
 H$^{13}$CN          &  3--2                                    &   25       & 259.01181       &   0.66 $\pm$   0.14  &    2.1          & 1.47 $\pm$ 0.75      & 191.1           & 0.21 &    \nodata \\
 c-C$_3$H$_2$       &  3$_{2,1}$--2$_{1,2}$        &   18       & 244.22215       &            $<$0.4   &    \nodata      &             $<$1.2   &    \nodata      & 0.19 &    \nodata \\
 U244.556                 &  \nodata                          &   \nodata       & 244.5556       &   1.02 $\pm$   0.11  &    2.3          & 2.55 $\pm$ 0.68      & 192.0           & 0.19 &    (5) (6) \\
\tableline 
\textit{Position 3} & & & & & & & & \\
 CS                  &  5--4                                    &   35       & 244.93556       &   6.33 $\pm$   0.12  &    2.8          & 18.64 $\pm$ 0.83     & 192.5           & 0.18 &    \nodata \\
 SO                  &  6$_6$--5$_5$                            &   57       & 258.25583       &   4.46 $\pm$   0.14  &    2.7          & 12.93 $\pm$ 0.92     & 192.4           & 0.19 &    \nodata \\
 H$^{13}$CO$^+$      &  3--2                                    &   25       & 260.25533       &   0.52 $\pm$   0.11  &    2.5          & 1.39 $\pm$ 0.63      & 192.4           & 0.19 &    \nodata \\
 CH$_3$OH            &  5$_{0}$--4$_{0}$ A$^+$                  &   35       & 241.79143       &   1.40 $\pm$   0.13  &    2.4          & 3.62 $\pm$ 0.80      & 192.5           & 0.21 &    \nodata \\
 CH$_3$OH            &  5$_{-1}$--4$_{-1}$ E                    &   40       & 241.76722       &   0.90 $\pm$   0.13  &    2.6          & 2.52 $\pm$ 0.82      & 192.1           & 0.21 &    \nodata \\
 CH$_3$OH            &  5$_{0}$--4$_{0}$ E                      &   48       & 241.70022       &             $<$0.4   &    \nodata      &             $<$1.3   &    \nodata      & 0.21 &    \nodata \\
 CH$_3$OH            &  5$_{1}$--4$_{1}$ A$^-$                  &   50       & 243.91583       &             $<$0.4   &    \nodata      &             $<$1.2   &    \nodata      & 0.18 &    \nodata \\
 CH$_3$OH            &  5$_{1}$--4$_{1}$ E                      &   56       & 241.87907       &             $<$0.4   &    \nodata      &             $<$1.3   &    \nodata      & 0.21 &    \nodata \\
 CH$_3$OH            &  5$_{\pm2}$--4$_{\pm2}$ E     &   61       & 241.90416       &             $<$0.4   &    \nodata      &             $<$1.3   &    \nodata      & 0.21 &    (4)  \\
 CH$_3$OH            &  5$_{2}$--4$_{2}$ A$^+$                  &   73       & 241.88771       &             $<$0.4   &    \nodata      &             $<$1.3   &    \nodata      & 0.21 &    \nodata \\
 SO$_2$              &  5$_{2,4}$--4$_{1,3}$                    &   24       & 241.61580       &   0.97 $\pm$   0.12  &    2.0          & 2.10 $\pm$ 0.72      & 192.2           & 0.21 &    \nodata \\
 C$^{33}$S           &  5--4                                    &   35       & 242.91361       &             $<$0.4   &    \nodata      &             $<$1.3   &    \nodata      & 0.21 &    \nodata  \\ 
 $^{33}$SO              &  6$_{7}$--5$_{6}$          &   47       & 259.28403       &             $<$0.4   &    \nodata      &             $<$1.2   &    \nodata      & 0.19 &    (1) \\
 c-C$_3$H$_2$       &  3$_{2,1}$--2$_{1,2}$        &   18       & 244.22215       &            $<$0.4   &    \nodata      &             $<$1.2   &    \nodata      & 0.18 &    \nodata \\
%%%%%
\enddata
\tablecomments{
\scriptsize{
Uncertainties and upper limits are of the 2$\sigma$ level and do not include systematic errors due to baseline subtraction. 
Upper limits are estimated assuming $\Delta$$V$ = 3 km s$^{-1}$. 
Notes: 
(1) Two hyperfine components (F = 15/2--13/2 and F = 17/2--15/2) are blended. 
(2) Tentative detection. 
(3) The integrated intensity is calculated by directly integrating the spectrum between 190 km s$^{-1}$ and 194 km s$^{-1}$. 
(4) Two CH$_3$OH lines (5$_{-2}$--4$_{-2}$ E and 5$_{2}$--4$_{2}$ E) are blended. 
(5) Tentatively attributed to HDS ($1_{0,1}-0_{0,0}$) (see Section \ref{sec_uline}). 
(6) Assuming $V_{\mathrm{LSR}}$ = 192.0 km s$^{-1}$. 
}}
\end{deluxetable*}
%%%%%%%%%%

%%%%%%%%%%
\begin{figure*}[tp]
\begin{center}
\includegraphics[width=18.5cm]{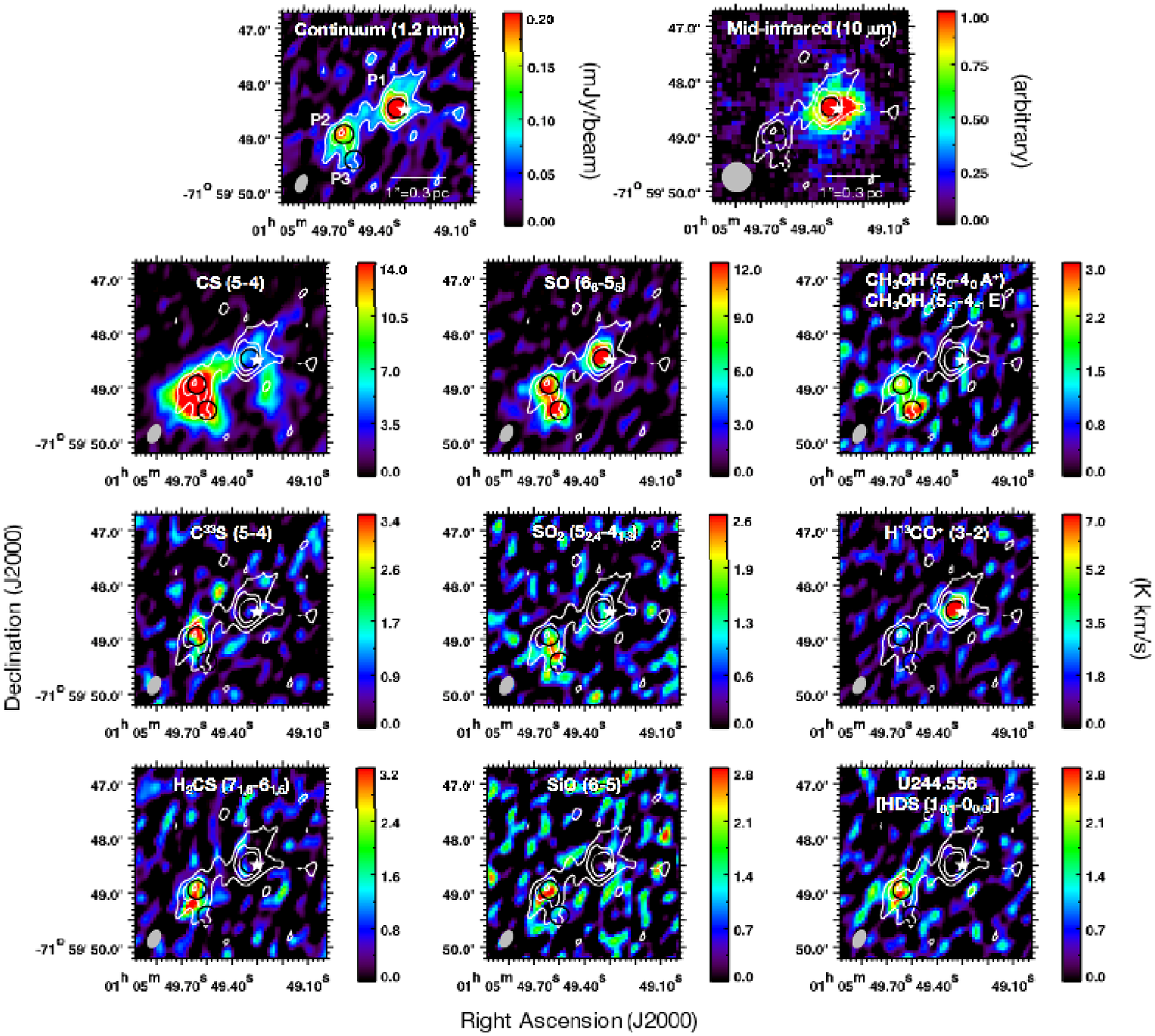}
\caption{
Flux distributions of the ALMA 1.2 mm continuum (top left) and the Gemini/T-ReCS mid-infrared 10 $\mu$m emission (top right), and integrated intensity distributions of CS, SO, CH$_3$OH, C$^{33}$S, SO$_2$, H$^{13}$CO$^{+}$, H$_2$CS, SiO, and U244.556 (tentatively attributed to HDS). 
Contours represent the 1.2 mm continuum distribution and the contour levels are 3$\sigma$, 6$\sigma$, and 10$\sigma$ of the rms noise. 
The synthesized beam size (0.11 pc $\times$ 0.07 pc at the SMC) is shown by the gray filled ellipse in each panel except for the 10 $\mu$m data, whose circle represents an FWHM of a standard star's point spread function. 
The spectra discussed in the text are extracted from the three positions (P1, P2, and P3) indicated by black open circles in each panel. 
The white filled star represents the position of an embedded high-mass YSO identified by infrared observations. 
North is up, and east is to the left. 
}
\label{images}
\end{center}
\end{figure*}
%%%%%%%%%%

%%%%%%%%%%
\begin{figure*}[tbp]
\begin{center}
\includegraphics[width=11cm]{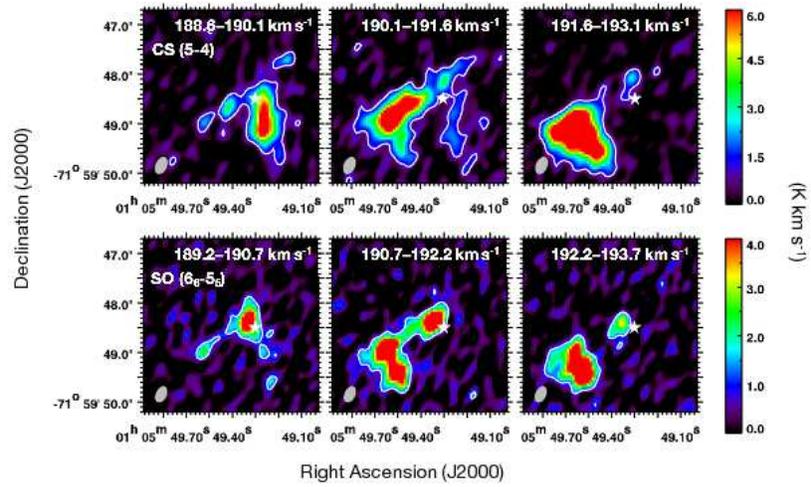}
\caption{
Channel maps for CS (up) and SO (bottom) at three velocity ranges (CS: 188.6-190.1, 190.1-191.6, 191.6-193.1 km s$^{-1}$; SO: 189.2-190.7, 190.7-192.2, 192.2-193.7 km s$^{-1}$). 
The plotted symbols are the same as in Figure \ref{images}. 
The contours represent the regions above 4$\sigma$ of the rms noise, and the filled star is the position of an embedded high-mass YSO. 
A velocity gradient from the YSO core to the East core is seen in the figure. 
}
\label{images_velo}
\end{center}
\end{figure*}
%%%%%%%%%%

\subsection{Synthesized images} \label{sec_img} 
Figure \ref{images} shows synthesized images of continuum and molecular emission lines observed toward the target region. 
The images are constructed by integrating spectral data in the velocity range where the emission is seen. 
For CH$_3$OH, two lines with similar spectroscopic parameters (5$_{0}$--4$_{0}$ A$^+$ and 5$_{-1}$--4$_{-1}$ E) are added together to increase the signal-to-noise
ratio (S/N). 
The H$^{13}$CN map is not shown here because the line is too weak to derive the distribution. 
In addition to the submillimeter data, a high spatial resolution mid-infrared 10 $\mu$m image obtained with Gemini/T-ReCS is shown in the figure. 

The distribution of continuum emission shows two peaks; one is associated with an embedded high-mass YSO (P1), and another is located at the east side of the YSO (P2) separated by about 1$\arcsec$ (0.3 pc). 
The latter is not seen in the mid-infrared image. 
The near-infrared image presented in \citet{War17} shows compact emission centered around P1, which is similar to our mid-infrared image. 
Note that the criterion for the continuum detection is the 3$\sigma$ significance level, and the S/Ns at P1, P2, and P3 are 14, 8, and 3, respectively. 
It seems that the YSO's position is slightly offset from the center of P1, but the separation is smaller than half of the beam size, and the positional accuracy of the infrared source is about 0.3$\arcsec$, thus we regard that the YSO and P1 are spatially coincident within the accuracy of the present dataset. 
Hereafter, the molecular cloud core at the YSO's position is referred to as \textit{the YSO core} and the one on the east side as \textit{the East core}. 
The size of the continuum emission region (FWHM of a two-dimensional Gaussian fit) is about 0.5--0.6$\arcsec$ (0.15--0.18 pc) at both P1 and P2, which are slightly larger than the beam size. 
P3 is located almost at the edge of the continuum emitting region in the East core. 

Molecular gas shows different distributions depending on species. 
SO is seen at all the positions (P1, P2, P3), while CS is concentrated in the East core (P2, P3) and the emission is weak at the YSO core. 
The distribution of CS is more extended compared with other lines and continuum, and the size of the emission region is about 0.8$\arcsec$ (0.24 pc) centered near P2. 
The CS emission is relatively weak at P1 and a peculiar structure elongated in the north-south direction is seen at the right side of P1. 
In contrast to CS, H$^{13}$CO$^{+}$ is concentrated at the YSO core and the source size is comparable to the beam size. 
CH$_3$OH is distributed at both P2 and P3, but the emission is slightly stronger at P3. 
The size of the CH$_3$OH emission region around P3 is about 0.4$\arcsec$ (0.12 pc), which is almost the same as the beam size. 
SO$_2$ is seen at both P2 and P3 although the S/N is not very good. 
The distribution of H$_2$CS seems to be slightly elongated toward the southeast direction from P2. 
Other lines, including C$^{33}$S, SiO, and U244.556, are distributed around P2. 
Note that all of the detected emission distributions are compact as compared to the maximum recoverable angular scale of the present observations ($\sim$3$\arcsec$), and thus the compact emission from the observed region is almost covered with the present array configuration. 

A bridge-like structure, which connects the YSO core and the East core, is seen in the continuum, CS, and SO images. 
Figure \ref{images_velo} shows channel maps for the CS and SO emission lines. 
A velocity gradient is clearly seen in the figure; the YSO core is relatively blue-shifted while the East core is red-shifted, and the bridge region shows an intermediate velocity. 
The velocity difference among those cores is at most 4 km s$^{-1}$. 
The continuous velocity structure from the YSO core to the bridge, and to the East core, suggests that these regions are dynamically associated.

\subsection{Column densities, gas temperatures, and molecular abundances} \label{sec_ntx}
\subsubsection{Rotation diagram analysis for CH$_3$OH} \label{sec_rd}
\begin{figure}[tp]
\begin{center}
\includegraphics[width=7.5cm]{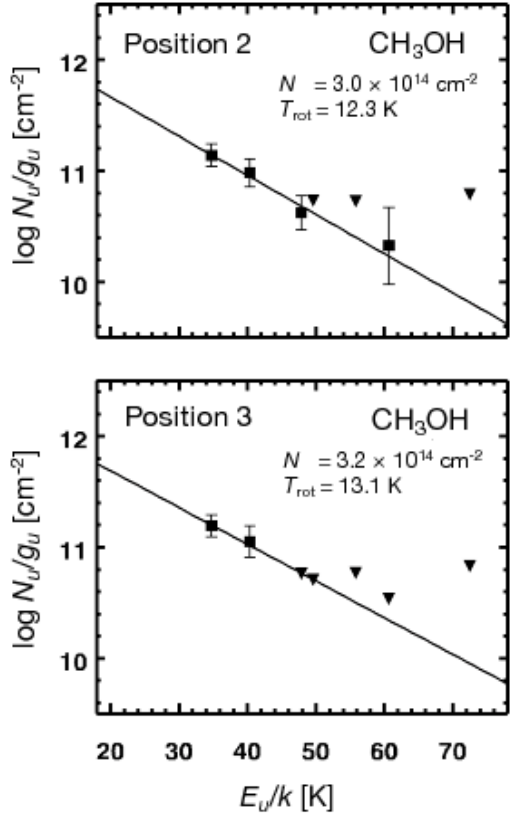}
\caption{
Rotation diagrams for CH$_3$OH lines at Position 2 (top) and 3 (bottom). 
Upper-limit points are shown by the downward-pointing triangles. 
The solid lines represent the fitted straight line. 
Derived column densities and rotation temperatures are indicated in each panel. 
Note that A- and E-state CH$_3$OH molecules are fitted simultaneously. 
See Section \ref{sec_rd} for details. 
}
\label{rd}
\end{center}
\end{figure}

We perform the rotation diagram analysis for CH$_3$OH because we detect multiple lines with different excitation energies (Figure \ref{rd}). 
An optically thin condition and the local thermodynamic equilibrium (LTE) are assumed here. 
We use the following formulae based on the standard treatment of the rotation diagram analysis \citep[e.g., ][]{Sut95, Gol99}:  
\begin{equation}
\frac{ N_{u} }{ g_{u} } = \frac{ 3 k \int T_{\mathrm{b}} dV }{ 8 \pi^{3} \nu S \mu^{2} }, \label{Eq_rd1} \\ 
\end{equation}
and
\begin{equation}
\log \left(\frac{ N_{u} }{ g_{u} } \right) = - \left(\frac {\log e}{T_{\mathrm{rot}}} \right) \left(\frac{E_{u}}{k} \right) + \log \left(\frac{N}{Q(T_{\mathrm{rot}})} \right),  \label{Eq_rd2}
\end{equation}
where $N_{u}$ is a column density of molecules in the upper energy level, $g_{u}$ is the degeneracy of the upper level, $k$ is the Boltzmann constant, $\int T_{\mathrm{b}} dV$ is the integrated intensity as estimated from the observations, $\nu$ is the transition frequency, $S$ is the line strength, $\mu$ is the dipole moment, $T_{\mathrm{rot}}$ is the rotational temperature, $E_{u}$ is the upper state energy, $N$ is a total column density, and $Q(T_{\mathrm{rot}})$ is the partition function at $T_{\mathrm{rot}}$. 
All the spectroscopic parameters required in the analysis are extracted from the CDMS database. 

Derived column densities and rotational temperatures are 
$N$ = (3.0 $^{+5.1}_{-1.9}$) $\times$ 10$^{14}$ cm$^{-2}$ and 
$T_{\mathrm{rot}}$ = 12.3 $^{+5.3}_{-2.9}$ K for Position 2, 
and 
$N$ = (3.2 $^{+4.5}_{-1.9}$) $\times$ 10$^{14}$ cm$^{-2}$ and 
$T_{\mathrm{rot}}$ = 13.1 $^{+4.9}_{-2.8}$ K for Position 3 (2$\sigma$ uncertainties). 
Note that the measured intensity of the CH$_3$OH line at 241.90416 GHz is divided by two, because it is a blend of two components (5$_{-2}$--4$_{-2}$ E and 5$_{2}$--4$_{2}$ E) with similar spectroscopic properties. 

In the above analysis, the data points for two marginal CH$_3$OH detections (5$_{0}$--4$_{0}$ E and 5$_{\pm2}$--4$_{\pm2}$ E, $Eu$ = 48 K and 61 K) are included in the straight-line fit for P2. 
For P3, two upper limit points for CH$_3$OH(5$_{0}$--4$_{0}$ E) and (5$_{1}$--4$_{1}$ A$^-$) ($Eu$ = 50 K) are included in the fit. 
If we neglect these marginal detections or upper limits in the fit, we obtain higher temperatures and slightly lower column densities; 
for P2, $T_{\mathrm{rot}}$ = 15.6 K and $N$ = 2.5 $\times$ 10$^{14}$ cm$^{-2}$, 
and 
for P3, $T_{\mathrm{rot}}$ = 17.3 K and $N$ = 2.8 $\times$ 10$^{14}$ cm$^{-2}$.   
However, these fits are incompatible with upper limits, and the fitted line deviates upward from the points. 
Thus, we take account of the above two detections in P2 and the two upper limit data in P3 to better fit the observation data. 

Note that the presence of warm gas components ($\gtrsim$50 K) within the core can not be ruled out from the limited frequency coverage of the present observations. 
Future observations in higher frequency regions are thus desirable. 

%%%%%%%%%%
%%%%%%%%%%

\subsubsection{Column density of H$_2$, dust extinction, and mass} \label{sec_h2} 
A column density of molecular hydrogen ($N_{\mathrm{H_2}}$) is estimated from the dust continuum data obtained in this work. 
The measured continuum brightnesses are 0.30 mJy/beam for P1, 0.16 mJy/beam for P2, and 0.06 mJy/beam for P3. 
The following equation is used to calculate $N_{\mathrm{H_2}}$ based on the standard treatment of optically thin dust emission \citep[e.g.,][]{ST16b}: 
\begin{equation}
N_{\mathrm{H_2}} = \frac{F_{\nu} / \Omega}{2 \kappa_{\nu} B_{\nu}(T_{d}) Z \mu m_{\mathrm{H}}} \label{Eq_h2}, 
\end{equation}
where $F_{\nu}/\Omega$ is the continuum flux density per beam solid angle as estimated from the observations, $\kappa_{\nu}$ is the mass absorption coefficient of dust grains coated by thin ice mantles as taken from \citet[][here we use 1.1 cm$^2$ g$^{-1}$]{Oss94}, $T_{d}$ is the dust temperature, $B_{\nu}(T_{d})$ is the Planck function, $Z$ is the dust-to-gas mass ratio, $\mu$ is the mean atomic mass per hydrogen \citep[1.41, according to][]{Cox00}, and $m_{\mathrm{H}}$ is the hydrogen mass. 
We assume that the dust-to-gas mass ratio in the SMC is lower than the typical Galactic value of 0.008 by a factor of five according to the metallicity. 
This assumption would be valid in a high density regime although dust-to-gas mass ratios in SMC diffuse clouds are higher than the value adopted here \citep[e.g.,][]{Rom17}. 

The dust temperature at P2 and P3 is assumed to be 12 K and 13 K according to the rotation temperatures of CH$_3$OH under the assumption of the LTE. 
In sub-thermal conditions, however, the kinetic temperature of gas is often higher than excitation temperatures. 
If we assume a slightly elevated dust temperature, i.e. $T_{d}$ = 20 K, the derived H$_2$ column densities decrease by a factor of two, and thus fractional molecular abundances are doubled. 
We should keep in mind that derived molecular abundances entail the systematic uncertainty depending on the assumption of dust temperatures. 
The dust temperature at P1 is unknown. 
We thus vary the dust temperature from 10 K to 20 K to derive the range of possible H$_2$ column densities. 
Assumption of higher dust temperatures ($T_{d}$ $>$ 23 K) for P1 leads to a low $A_{V}$ value ($A_{V}$ $<$ 8 mag), which is unrealistic as extinction towards a deeply embedded high-mass YSO, and also inconsistent with the previous $A_{V}$ estimate described below. 

The derived H$_2$ column densities are shown in Table \ref{tab_N} and used to calculate fractional abundances of other molecular species. 
The total gas mass contained in each core is estimated to be about 80 M$_{\sun}$ for the YSO core and about 50 M$_{\sun}$ for the East core (assuming $T_{d}$ = 10 K for both cores). 

We also estimate visual extinction ($A_{V}$) using an $N_{\mathrm{H_2}}$/$A_{V}$ conversion factor. 
\citet{Mar89} reported $N_{\mathrm{H}}$/$E(B-V)$ = 4.6 $\times$ 10$^{22}$ cm$^{-2}$ mag$^{-1}$ for the SMC. 
Using $A_{V}$/$E(B-V)$ $\sim$4 for dense clouds \citep{Whi01b}, we obtain $N_{\mathrm{H_2}}$/$A_{V}$ = 5.8 $\times$ 10$^{21}$ cm$^{-2}$ mag$^{-1}$, where all the hydrogen atoms are assumed to be in H$_2$. 
A similar ratio is obtained when the Galactic canonical $N_{\mathrm{H}}$/$A_{V}$ value (= 1.9 $\times$ 10$^{21}$ cm$^{-2}$ mag$^{-1}$) is scaled by the metallicity of the SMC. 
Using the above $N_{\mathrm{H_2}}$/$A_{V}$ ratio, we estimate $A_{V}$ $\sim$10--30 mag at P1, $\sim$10 mag at P2, and $\sim$4 mag at P3. 
\citet{War17} estimate $A_{V}$ = 16 $\pm$ 8 mag for the YSO at P1 using the ratio of near-infrared H$_2$ emission lines; the value agrees with our $A_{V}$ estimate that is based on the submillimeter continuum. 
Note that these are rough estimates since the $N_{\mathrm{H_2}}$/$A_{V}$ ratio is highly uncertain for dense clouds in the SMC.

\subsubsection{Column densities of other molecules} \label{sec_n} 
Column densities of molecular species other than CH$_3$OH and H$_2$ are estimated by solving Equation \ref{Eq_rd2} for $N$. 
As well as the dust continuum analysis described in Section \ref{sec_h2}, the gas temperature at P2 and P3 is assumed to be 12 K and 13 K, while the gas temperature at P1 is varied between 10 K and 20 K. 

The CS(5--4) and SO(6$_6$--5$_5$) lines are often optically thick in star-forming regions. 
For CS at P1 and P3, we use the C$^{32}$S data to derive a lower limit and the C$^{33}$S non-detection data for an upper limit to estimate a possible range of column densities. 
At P2, the CS column density is derived only from the C$^{33}$S line. 
Similarly, for SO, we use the $^{32}$SO data to derive a lower limit and the $^{33}$SO non-detection data for an upper limit to estimate a possible range of column densities at all positions. 
We here assume $^{32}$S/$^{33}$S = 40, based on the $^{33}$SO$_2$ observations towards a hot core in the LMC \citep{ST16b}. 
Column densities of HCO$^+$ and HCN are estimated from H$^{13}$CO$^+$ and H$^{13}$CN data, respectively. 
We here assume $^{12}$C/$^{13}$C = 50 according to HCO$^+$/H$^{13}$CO$^+$ observations of the N27 region in the SMC \citep{Hei99}. 

Estimated column densities and fractional abundances at each position are summarized in Table \ref{tab_N}. 

%%%%%%%%%%
\begin{deluxetable*}{ l c c c c c c }
\tablecaption{Column densities and fractional abundances \label{tab_N}}
\tablewidth{0pt}
\tabletypesize{\small} %preprint
\tablehead{
\colhead{ }                & \multicolumn{2}{c}{Position 1\tablenotemark{a}}  &  \multicolumn{2}{c}{Position 2\tablenotemark{b}} & \multicolumn{2}{c}{Position 3\tablenotemark{b}}  \\
                                                                                            \cline{2-7}                                                                             \\
\colhead{Molecule}   & \colhead{$N$}              &  \colhead{$N$/$N_{\mathrm{H_2}}$}  &  \colhead{$N$}            &  \colhead{$N$/$N_{\mathrm{H_2}}$}  &  \colhead{$N$}             &  \colhead{$N$/$N_{\mathrm{H_2}}$}  \\
\colhead{ }                &   \colhead{(cm$^{-2}$)} &  \colhead{ }                                         &   \colhead{(cm$^{-2}$)} &  \colhead{ }                                          &   \colhead{(cm$^{-2}$)} &  \colhead{ }
}
\startdata 
H$_2$\tablenotemark{c}         &   (0.6--1.6) $\times$ 10$^{23}$       &  \nodata                                             &  6.3  $\times$ 10$^{22}$                            &  \nodata                                                     &  2.1  $\times$ 10$^{22}$                              &  \nodata  \\
CH$_3$OH\tablenotemark{d} & ($<$1.1--1.8) $\times$ 10$^{14}$   & $<$2 $\times$ 10$^{-9}$                  & (3.0 $^{+5.1}_{-1.9}$) $\times$ 10$^{14}$ & (4.7 $^{+8.0}_{-3.0}$) $\times$ 10$^{-9}$ & (3.2 $^{+4.5}_{-1.9}$) $\times$ 10$^{14}$  & (1.5$^{+2.2}_{-0.9}$) $\times$ 10$^{-8}$   \\
CS\tablenotemark{e}            &  (0.2--5.4) $\times$ 10$^{14}$        &  (1.8 $\pm$ 1.5) $\times$ 10$^{-9}$  &  (1.3 $\pm$ 0.3) $\times$ 10$^{15}$          &  (2.0 $\pm$ 0.4) $\times$ 10$^{-8}$         &  (1.1--3.3) $\times$ 10$^{14}$                    & (1.0 $\pm$ 0.5) $\times$ 10$^{-8}$    \\
SO\tablenotemark{f}              &  (0.5--6.4) $\times$ 10$^{15}$       & (2.4 $\pm$ 1.6) $\times$ 10$^{-8}$    &  (1.4--4.0) $\times$ 10$^{15}$                    & (4.3 $\pm$ 2.1) $\times$ 10$^{-8}$           & (1.1--3.0) $\times$ 10$^{15}$                     &  (9.5 $\pm$ 4.5) $\times$ 10$^{-8}$   \\
SO$_2$                                  &  ($<$5.2--6.0) $\times$ 10$^{13}$  & $<$9 $\times$ 10$^{-10}$                 & (9.9 $\pm$ 6.1) $\times$ 10$^{13}$           & (1.6 $\pm$ 1.0) $\times$ 10$^{-9}$          & (8.4 $\pm$ 2.9) $\times$ 10$^{13}$            & (4.0 $\pm$ 1.4) $\times$ 10$^{-9}$  \\
H$_2$CS                               &  ($<$0.3--1.8) $\times$ 10$^{14}$   &  $<$1 $\times$ 10$^{-9}$                  &  (2.4 $\pm$ 0.5) $\times$ 10$^{14}$          &  (3.8 $\pm$ 0.8) $\times$ 10$^{-9}$         &  $<$6.9 $\times$ 10$^{13}$                        &  $<$3 $\times$ 10$^{-9}$    \\
HCN\tablenotemark{g}          &  ($<$5.1--9.1) $\times$ 10$^{13}$   &  $<$9 $\times$ 10$^{-10}$                &  (9.6 $\pm$ 4.4) $\times$ 10$^{13}$          &  (1.4 $\pm$ 0.7) $\times$ 10$^{-9}$          &  $<$6.5 $\times$ 10$^{13}$                        &  $<$3 $\times$ 10$^{-9}$    \\
HCO$^+$\tablenotemark{g}  &  (1.7--4.1) $\times$ 10$^{14}$         &  (2.9 $\pm$ 1.0) $\times$ 10$^{-9}$  &  (7.9 $\pm$ 3.3) $\times$ 10$^{13}$          & (1.3 $\pm$ 0.5) $\times$ 10$^{-9}$          &  (4.4 $\pm$ 2.0) $\times$ 10$^{13}$           &  (2.1 $\pm$ 1.0) $\times$ 10$^{-9}$    \\
c-C$_3$H$_2$                      &  ($<$5.5--6.3) $\times$ 10$^{13}$   &  $<$1 $\times$ 10$^{-9}$                  &  $<$5.3 $\times$ 10$^{13}$                       &  $<$9 $\times$ 10$^{-10}$                       &  $<$5.1 $\times$ 10$^{13}$                        &  $<$2 $\times$ 10$^{-9}$     \\ 
SiO                                        &  ($<$0.2--1.1) $\times$ 10$^{13}$   &  $<$7 $\times$ 10$^{-11}$                &  (1.3 $\pm$ 0.6) $\times$ 10$^{13}$          &  (2.1 $\pm$ 0.9) $\times$ 10$^{-10}$       &  $<$7.9 $\times$ 10$^{12}$                        &  $<$4 $\times$ 10$^{-10}$     \\ 
\enddata
\tablecomments{
Uncertainties and upper limits are of the 2$\sigma$ level and do not include systematic errors due to adopted spectroscopic constants. 
$^a$$T_{\mathrm{rot}}$ varied between 10 K and 20 K. 
$^b$$T_{\mathrm{rot}}$ assumed to be 12 K for P2 and 13 K for P3 based on the rotation analysis of CH$_3$OH lines. 
$^c$Estimated from dust continuum. 
$^d$Estimated based on the rotation analysis. 
$^e$The lower limit and the upper limit are estimated from the CS and C$^{33}$S data (assuming $^{32}$S/$^{33}$S = 40). For Position 2, only the C$^{33}$S data are used. 
$^f$The lower limit and the upper limit are estimated from the SO and $^{33}$SO data using the above isotope ratio. 
$^g$Estimated from the isotopologue assuming  $^{12}$C/$^{13}$C = 50. 
See Section \ref{sec_ntx} for details. 
} 
\end{deluxetable*}
%%%%%%%%%%

\section{Discussion} \label{sec_disc} 
\subsection{Physical properties of two dense cores} \label{sec_reg} 
The present ALMA observations resolved a single infrared source into two molecular cloud cores (the YSO core and the East core). 
They are physically associated but show different chemical properties. 
We first discuss physical properties of these cores in the following section. 

\subsubsection{The YSO core} \label{sec_reg1} 
\begin{figure}[tp]
\begin{center}
\includegraphics[width=8.0cm]{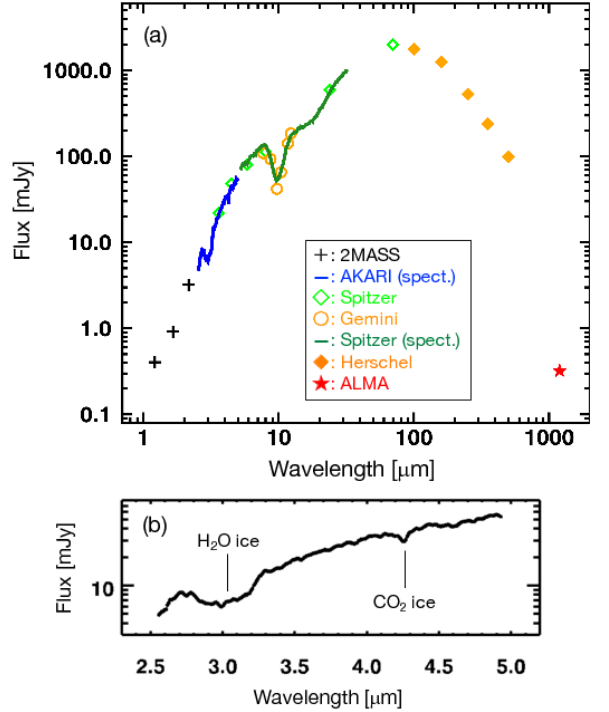}
\caption{
(a): The SED of a high-mass YSO located at Position 1. 
The plotted data are based on 2MASS JHK$_\mathrm{s}$ photometry (pluses, black), {\it AKARI}/IRC spectroscopy (solid line, blue), {\it Spitzer}/IRAC and MIPS photometry (open diamonds, light green), Gemini/T-ReCS narrow-band photometry (open circles, orange), {\it Spitzer}/MIPS spectroscopy (solid line, green), {\it Herschel}/PACS and SPIRE photometry (filled diamonds, orange), and ALMA 1200 $\mu$m continuum measurement obtained in this work (filled star,red). 
(b): Zoom-in of the 2--5 $\mu$m spectrum obtained by {\it AKARI}. 
The positions of the 3.05 $\mu$m H$_2$O ice band and the 4.27 $\mu$m CO$_2$ ice band are labeled. 
See Section \ref{sec_disc} for more details. 
}
\label{sed}
\end{center}
\end{figure}

The molecular cloud core including Position 1 is associated with an embedded high-mass YSO that is bright in infrared and radio continuum. 
A spectral energy distribution (SED) of the YSO is shown in Figure \ref{sed} \citep[data are collected from available databases and literatures, including][]{Skr06, Mei06, Mei13, Kem10}. 
The photometric point at 1200 $\mu$m is estimated from the present ALMA data. 
The 2--5 $\mu$m spectrum is based on our {\it AKARI} data, and the five narrow-band photometric points around 10 $\mu$m are based on our Gemini/T-ReCS data (see Section \ref{sec_ir} for details of these data). 
The YSO's SED peaks in the far-infrared region ($\sim$50-70 $\mu$m), suggesting that the source is at an early evolutionary stage. 
The bolometric luminosity of the YSO is estimated to be 2 $\times$ 10$^4$ L$_{\sun}$ by integrating the SED from 1 $\mu$m to 1200 $\mu$m. 
\citet{Oli13} reported a luminosity of 2.3 $\times$ 10$^4$ L$_{\sun}$ for the YSO, with which our estimate is consistent. 
The high luminosity is consistent with the presence of a high-mass YSO. 

%%%%%%%%%%
%%%%%%%%%%

Absorption bands due to the H$_2$O ice, the CO$_2$ ice, and the silicate dust are seen in the 2--20 $\mu$m spectrum of the YSO (Fig. \ref{sed}), as previously reported in \citet{Oli11,Oli13}. 
The beam size of the \textit{AKARI} data, which shows the ice absorption bands, is much larger than those of ALMA and Gemini data. 
However, because ices are observed with absorption spectroscopy, the absence of an infrared continuum source at P2 and P3 as shown in the infrared image indicates that the observed ice absorption bands originate mainly from the infrared point source associated with P1. 
Further analysis on ice absorption bands is described in Section \ref{sec_chemdif}. 
Previous studies detected compact emission due to hydrogen recombination lines and shocked H$_2$ lines associated with the YSO \citep{Oli13,War17}. 
On the other hand, emission features due to fine-structure lines from ionized metals and polycyclic aromatic hydrocarbon (PAH) bands are absent or weak in the infrared spectrum of the YSO. 
The above spectral characteristics suggest that the YSO is still embedded in cold dust and ices, and a prominent \ion{H}{2} region is not formed yet. 

A number density of molecular hydrogen at P1 is estimated to be $n_{\mathrm{H_2}}$ $\sim$3--8 $\times$ 10$^5$ cm$^{-3}$ (assuming a source diameter of 0.1 pc and a spherical distribution of gas) and the $A_{V}$ in the line of sight is $\sim$10--30 mag. 
This suggests that the YSO core contains dense molecular gas that is well shielded from the external radiation field. 
Given the embedded nature of the source and the absence of a prominent \ion{H}{2} region, the radiation from the central protostar would only have a limited effect on the bulk of molecular gas within the core.

\subsubsection{The East core} \label{sec_reg2} 
The molecular cloud core including Position 2 and 3 is not associated with an infrared source. 
The gas density (assuming the source diameter of 0.1 pc and the spherical distribution) and the dust extinction are estimated to be $n_{\mathrm{H_2}}$ = 3 $\times$ 10$^5$ cm$^{-3}$ and $A_{V}$ $\sim$10 mag at P2, while $n_{\mathrm{H_2}}$ = 1 $\times$ 10$^5$ cm$^{-3}$ and $A_{V}$ $\sim$4 mag at P3. 
The temperature of molecular gas is $\sim$10 K as estimated from CH$_3$OH lines (see Section \ref{sec_rd}). 
These properties suggest that the East core contains cold and dense molecular gas. 

The lack of infrared sources suggests that a high-mass YSO is not present in the East core except for extremely young and cold ones that are elusive even in infrared. 
However, the possible presence of low-mass YSOs in the East core cannot be ruled out from the current infrared data because of the sensitivity limit. 
A single low-mass YSO would be too faint to be detected by the present submillimeter observations since the size of a protostellar core is usually small \citep[$\lesssim$0.01 pc, e.g.,][]{vD95}, but an unresolved cluster of low-mass YSOs may produce a detectable amount of molecular line emission. 
A massive starless cloud such as an infrared dark cloud is also a possible candidate for the East core. 

The low temperature of CH$_3$OH gas, as well as the lack of a bright infrared source in the East core, suggests that the presence of a prominent hot core would be less likely. 
Cold CH$_3$OH gas ($\sim$5 K) has been detected in Galactic low-mass starless cores, but the observed line width is typically 0.1--0.5 km s$^{-1}$, which is much narrower than the line width measured for the East core \citep{Taf06, Som15}. 

The above discussion suggests that the East core would harbor (i) a massive starless core, or (ii) an embedded high-mass YSO(s) before the emergence of infrared emission, or (iii) a cluster of low-mass embedded YSOs.

\subsection{Chemical differentiation} \label{sec_chemdif} 
\begin{figure*}[tp]
\begin{center}
\includegraphics[width=17cm]{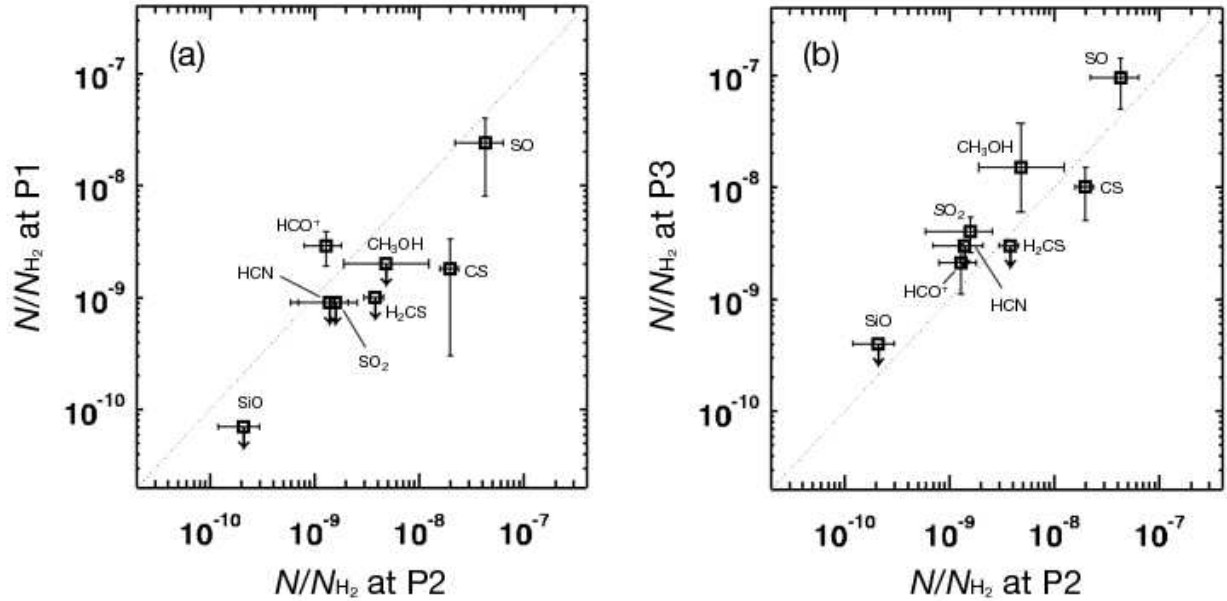}
\caption{
Comparisons of molecular gas abundances at different positions; (a) P1 vs. P2 and (b) P3 vs. P2. 
Upper and lower limits are indicated by arrows. 
The dotted lines represent an abundance ratio of 1. 
Data of c-C$_3$H$_2$ are not plotted since it is not detected at all positions. 
See Section \ref{sec_chemdif} for a detailed discussion. 
}
\label{comparison}
\end{center}
\end{figure*}

Chemical compositions of the YSO core and the East core, which are separated by about 0.3 pc, show different characteristics. 
Figure \ref{comparison} compares molecular abundances at P1, P2, and P3. 
The P1--P2 comparison shows that only HCO$^+$ is more abundant at P1, while CS, H$_2$CS, and SiO are significantly underabundant at P1. 
On the other hand, the P2--P3 comparison shows generally similar molecular abundances between P2 and P3. 
Note that CH$_3$OH and SO$_2$ are slightly overabundant at P3 compared to P2. 
The correlation coefficient for abundances of molecular species that are detected in both P2 and P3 (CH$_3$OH, CS, SO, SO$_2$, and HCO$^+$) is 0.89. 
The good correlation suggests a similar molecular abundances within the East core. 
These results imply a chemical differentiation between the YSO core and the East core. 

One possible interpretation of this chemical differentiation is the different degree of the adsorption of gas-phase species onto grain surfaces. 
As shown in Figure \ref{sed}(b), absorption bands due to the H$_2$O and CO$_2$ ices are detected in the line of sight toward the YSO located at P1. 
The estimated high $A_{V}$ toward P1 (10--30 mag) is actually sufficient for the formation of ice mantles. 
Using the \textit{AKARI} spectrum shown in the figure, ice column densities are estimated to be $N$(H$_2$O ice) = (17 $\pm$ 5) $\times$ 10$^{17}$ cm$^{-2}$ and $N$(CO$_2$ ice) = (1.1 $\pm$ 0.2) $\times$ 10$^{17}$ cm$^{-2}$ for the absorption bands at 3.05 $\mu$m and 4.27 $\mu$m \citep[derivation is based on the method presented in][]{ST16}. 
For the CH$_3$OH and CO ices, upper limits of $N$(CH$_3$OH ice) $\leq$ 4 $\times$ 10$^{17}$ cm$^{-2}$ and $N$(CO ice) $\leq$ 2 $\times$ 10$^{17}$ cm$^{-2}$ (2$\sigma$) are estimated for the absorption bands at 3.53 $\mu$m and 4.67 $\mu$m, respectively. 
\citet{Oli11,Oli13} have estimated ice column densities of the YSO to be $N$(H$_2$O ice) = (16.6 $\pm$ 0.7) $\times$ 10$^{17}$ cm$^{-2}$, $N$(CO$_2$ ice) = (1.0 $\pm$ 0.2) $\times$ 10$^{17}$ cm$^{-2}$, and $N$(CO ice) $\leq$ 0.45 $\times$ 10$^{17}$ cm$^{-2}$ using ground-based L-/M-band spectra and a \textit{Spitzer} mid-infrared spectrum. 
The present results of the \textit{AKARI} spectroscopy agree with their measurements. 

The total fractional abundance of elemental oxygen (w.r.t. H) in solid H$_2$O and CO$_2$ at P1 is estimated to be $\sim$1 $\times$ 10$^{-5}$, where we use the H$_2$ column density shown in Table \ref{tab_N}. 
Similarly, that for elemental carbon in ice mantles is estimated to be $\sim$6 $\times$ 10$^{-7}$. 
Elemental abundances of gas-phase oxygen and carbon in the SMC, after considering the depletion into dust grain material, are expected to be 3.7 $\times$ 10$^{-5}$ for oxygen and 1.5 $\times$ 10$^{-5}$ for carbon, according to the the SMC low-metal model of \citet{Ach16}. 
Although the gas-phase counterparts of H$_2$O and CO$_2$ are not observed in this work, the above discussion suggests that a non-negligible fraction of heavy elements, especially oxygen, are frozen in ice mantles at P1. 

Shock chemistry is another possible cause of the chemical differentiation between the YSO core and the East core. 
The detection of shock tracers such as SO$_2$ and SiO at P2/P3 implies a possible contribution of shock chemistry on the chemical compositions. 
\citet{War17} detected collisionally-excited near-infrared H$_2$ emission lines in IRAS 01042-7215, which suggests the presence of shocked gas in this region. 
The effect of shocks on the chemical compositions of the East core will be further discussed in Section \ref{sec_ch3oh_2}. 

Finally, we note that future high-sensitivity observations for ices in the East core will be important to further investigate the chemical differentiation in the observed region, since the estimated $A_{V}$ at P2 and P3 ($\sim$4--10 mag) is still plausible for the ice formation.

%%%%%%%%%%
%%%%%%%%%%

\subsection{Organic molecules in a low metallicity environment} \label{sec_ch3oh} 
The detection of CH$_3$OH in the SMC has a strong impact on our understanding of the formation of organic molecules in metal-poor environments. 
We here discuss abundances and possible origins of cold CH$_3$OH gas in the East core. 

\subsubsection{Abundance of CH$_3$OH} \label{sec_ch3oh_1} 
The fractional abundance of CH$_3$OH gas in the East core is (4.7 $^{+8.0}_{-3.0}$) $\times$ 10$^{-9}$ at P2 and 1.5$^{+2.2}_{-0.9}$ $\times$ 10$^{-8}$ at P3. 
As discussed in Section \ref{sec_reg2}, a possible Galactic counterpart for the East core will be cold and embedded sources such as infrared dark clouds or early-stage YSOs without a hot core region. 
Thus, the CH$_3$OH gas abundances estimated in this work should be compared with those of such Galactic sources. 
Typical abundances of CH$_3$OH gas in Galactic infrared dark clouds are reported to be several $\times$ 10$^{-9}$ \citep[see, e.g., Fig.3 in][]{Ger14}. 
Similar abundances are reported for Galactic embedded YSOs without hot cores/corinos \citep[e.g.,][]{vdT00, Wat12, Ger14}. 
Despite a factor of five lower metallicity in the SMC, the CH$_3$OH gas abundance at P2 is comparable to those of Galactic counterparts, and that of P3 is marginally higher than the typical Galactic value. 
If the CH$_3$OH gas abundance is simply regulated by the elemental abundances of carbon and oxygen, then the decreased CH$_3$OH abundance according to the SMC's metallicity is expected, but this is not observed in the present data. 
This would indicate an enhanced production or a suppressed destruction of gas-phase CH$_3$OH in the East core compared to Galactic similar sources. 

The CH$_3$OH gas abundance in the East core is comparable to those of two LMC hot cores reported in \citet{Sew18}, although the excitation temperature of CH$_3$OH in the East core is much lower than those of the two hot cores. 
Currently, a hot core is not yet detected in the SMC, thus future search for SMC hot cores will be important to further understand low-metallicity organic chemistry in warm and dense gas around a protostar. 

\subsubsection{Possible origin of CH$_3$OH in the East core} \label{sec_ch3oh_2} 
It is believed that CH$_3$OH is mainly formed on dust grains. 
Hydrogenation of CO on grain surfaces or energetic processes by photolysis/radiolysis of ice mantles are believed to be responsible for the formation of solid CH$_3$OH \citep[e.g.,][]{Ger01,Wat07}. 
The gas-phase CH$_3$OH is thus produced by desorption of solid CH$_3$OH from ices. 

The desorption mechanisms include (i) sublimation by heating, (ii) photodesorption by UV photons, (iii) sputtering by shocks \citep[e.g.,][]{Aot15}, and (iv) reactive desorption\footnote{Desorption triggered by liberation of reaction energy} \citep[e.g.,][]{Gar07}. 
The low gas temperature at P2 and P3 suggests that (i) would be less dominant in the East core, since the sublimation temperature of the CH$_3$OH ice is higher than $\sim$80 K \citep[e.g.,][]{Tie05}. 
A major source of the UV radiation in the East core would be the cosmic-ray-induced UV photons owing to the high $A_{V}$. 
However, the average cosmic-ray density in the SMC is reported to be lower than the solar neighborhood by a factor of $\sim$7 \citep{Abd10b}. 
This may lower the efficiency of (ii) compared to the Galactic dense clouds. 
Note that P3 is located close to the edge of the East core and that the estimated column density and extinction are lower than those at P1 and P2. 
Thus, the photodesorption of CH$_3$OH from ices could be enhanced at P3 compared to P1 and P2 owing to the less attenuated interstellar radiation field. 
Such a physical condition similar to photodissociation regions may contribute to the high abundance of CH$_3$OH gas at P3. 

A common shock tracer, SiO, is marginally detected at P2, and its line width is possibly broader than those of other emission lines observed in the East core. 
This suggests a possible contribution of the mechanism (iii) to the production of gas-phase CH$_3$OH. 
The energy source(s) responsible for triggering the shock chemistry are unknown; possible candidates are outflows from the high-mass YSO located at P1 or outflows from hidden embedded YSO(s) in the East core. 
Given the small velocity difference between the East core and the YSO core ($\sim$4 km/s, see Fig. \ref{images_velo}), the collision between the two cores and subsequent shock sputtering of ice mantles are less likely for the production of CH$_3$OH gas in the present source \citep[see shock sputtering models in ][]{Aot15, Miu17}. 
The presence of the dynamically continuous bridge-like structure seen in the continuum, CS, and SO lines would support the physical interaction between the YSO core and the East core (Section \ref{sec_img}). 
Further observations of intense shock/outflow tracers will help test the contribution of the mechanism (iii). 
Finally, the mechanism (iv) will be also possible, but its relative contribution should be carefully evaluated with the aid of astrochemical models. 

Note that the CH$_3$OH ice absorption band at 3.53 $\mu$m is not detected in the infrared spectrum of the YSO at P1. 
However, this does not necessarily mean the absence of solid CH$_3$OH in the YSO core, because the spectral resolution and sensitivity of the present infrared data are not sufficient to detect a weak absorption band of the CH$_3$OH ice.

\subsubsection{Comparison with observations of the LMC and the relevant chemical model} \label{sec_ch3oh_3} 
In the LMC, a low abundance of CH$_3$OH, compared to similar Galactic objects, is suggested for ices around embedded high-mass YSOs \citep{Sea11,ST16}, warm and dense gas around a hot molecular core \citep{ST16b}, and cold diffuse gas in molecular clouds \citep{Nis15}. 
The low CH$_3$OH abundance is also suggested for molecular clouds in another nearby low-metallicity galaxy, IC10 \citep{Nis16}. 
\citet{ST16} proposed the warm ice chemistry hypothesis to interpret the low abundance of CH$_3$OH in the LMC. 
The hypothesis argues that warm dust temperatures inhibit the hydrogenation of CO in LMC dense clouds, which contributes to the low CH$_3$OH abundances in both solid and gas phase. 
Astrochemical simulations on gas-ice chemistry dedicated to the LMC environment quantitatively characterize such temperature effects \citep{Ach15,Pau18}. 
On the other hand, \citet{Sew18} recently detected CH$_3$OH gas and its possible daughter species (CH$_3$OCH$_3$ and CH$_3$OCHO) from two hot cores in the LMC. 
They argue that abundances of these molecular species simply scale with the metallicity of the LMC. 
The results would suggest the large chemical variation among star-forming cores in the LMC and that the suppression of the CH$_3$OH production due to warm ice chemistry does not always dominate the organic chemistry at low metallicity. 

In the SMC, the metallicity is even lower than the LMC, and the interstellar radiation field is expected to be less attenuated. 
This would possibly raise dust temperatures in dense clouds, and the CH$_3$OH production could be suppressed if the dust temperature has a dominant effect on grain surface chemistry. 
However, the CH$_3$OH gas abundance in the East core is comparable with those of similar Galactic objects as shown in Section \ref{sec_ch3oh_1}. 
A decrease of the CH$_3$OH abundance, as expected from the inhibition of CO hydrogenation at elevated dust temperatures, is not seen in the present SMC source. 

\citet{ST16} argue that a high CO$_2$/H$_2$O ice column density ratio is another observational indicator of warm ice chemistry, because mobile CO on warm dust surfaces would efficiently react with OH to form CO$_2$. 
Previous observations reported that CO$_2$/H$_2$O ice ratios are about a factor of two larger in envelopes of high-mass YSOs in the LMC \citep[$\sim$30$\%$,][]{ST,ST10,ST16,Oli09,Sea11} as compared with those in Galactic high-mass YSOs \citep[$\sim$17$\%$,][]{Ger99}. 
The CO$_2$/H$_2$O ice ratio of the high-mass YSO at P1 is estimated to be 7$^{+4}_{-3}$$\%$ (Section \ref{sec_chemdif}), which is lower than the typical ratios for Galactic and LMC high-mass YSOs. 
Ice abundances in P2/P3 are currently unknown, but at least in P1, the observations suggest that the enhanced CO$_2$ ice abundance as observed in LMC YSOs is not clearly seen. 
The lack of the CO$_2$/H$_2$O ratio enhancement in the SMC is also suggested for a larger sample of embedded high-mass YSOs in the SMC \citep{Oli13}. 
\citet{Oli11} state that enhanced photodesorption of H$_2$O ice in the outer envelope of YSOs can explain the observed characteristics of the CO$_2$/H$_2$O ratios in the LMC/SMC, rather than the grain temperature effect mentioned above. 

The above discussions suggest that the chemical properties of dense cores in IRAS 01042-7215 show different characteristics compared with previously observed LMC sources, although statistical properties of dense cloud chemistry at the LMC's and SMC's metallicity are yet to be investigated. 

Numerical simulations of gas-ice chemistry dedicated to the SMC and further metal-poor environments are presented in the literature \citep{Ach16, Pau18, Cec18}.  
\citet{Ach16} show that low grain temperatures (lower than at least 20 K) and well-shielded conditions ($A_{V}$ larger than at least 5 mag) are necessary to attain a CH$_3$OH gas abundance higher than 10$^{-9}$ in SMC dense clouds (see their Fig. 3). 
The present observational results suggest that a cold, dense, and well-shielded region that is necessary for the production of CH$_3$OH actually exists even at the decreased metallicity of the SMC. 
Interestingly, the astrochemical simulations in \citet{Pau18} suggest the enhancement of the CH$_3$OH ice abundance relative to the CO ice in their low-metallicity models (particularly in their SMC models). 
According to their explanation, higher grain temperatures decrease the efficiency of the CO hydrogenation, because more CO is converted to CO$_2$ and less CH$_3$OH is produced, as expected from the warm ice chemistry hypothesis. 
On the other hand, they also argue that lower metallicities result in slower accretion rates of gaseous heavy elements, which relatively increases the efficiency of the CO hydrogenation, because this allows CO to reside on the surface for a longer time for the successive hydrogenation to form CH$_3$OH, before being embedded in the mantle phase and/or reacting with other heavy surface reactants such as OH. 
It is likely that the balance between the dust temperature and the abundance of heavy elements is one of the important determinants of the CO hydrogenation efficiency on grain surfaces. 
Since neither the CO ice nor the CH$_3$OH ice is detected towards the East core and other embedded sources in the SMC, future high-sensitivity observations need to statistically quantify abundances of solid CO and CH$_3$OH in the SMC. 
This may shed light on the reason for the abundant CH$_3$OH gas in the low-metallicity dense core observed in this work. 

We emphasize that it is still unclear whether the chemical composition observed in the East core of IRAS 01042-7215 represents common characteristics of cold and dense molecular clouds in metal-poor environments. 
Theoretical studies argue that chemical compositions of dense molecular clouds show a complex dependence on various local environmental factors, notably for species that are mainly produced on grain surfaces such as CH$_3$OH \citep{Ach15,Ach16,Pau18}. 
It may be possible that patchy distributions of the interstellar medium in an irregular galaxy cause localized and variable radiation environments, which can lead to diverse physical and chemical conditions of molecular clouds compared to our Galaxy. 
We obviously need to increase the relevant astrochemical data in future observations.

\subsection{Tentative Detection of HDS} \label{sec_uline} 
An unidentified line is detected at the rest-frame frequency of 244.5556 GHz (assuming $V_{\mathrm{LSR}}$ = 192.0 km s$^{-1}$) in the spectrum of Position 2. 
The spatial distribution of the emission is similar to those of C$^{33}$S, H$_2$CS, and SiO. 
The line width is consistent with the other lines detected in the East core. 

A possible candidate for the line is HDS ($1_{0,1}-0_{0,0}$), which is located at 244.55558 GHz and has the upper-state energy of 11.7 K according to the CDMS catalogue. 
The low upper-state energy suggests that this transition can be excited even in the low-temperature condition at P2. 
If the line is due to HDS ($1_{01}-0_{00}$), then the observed line strength corresponds to $N$(HDS) = (5 $\pm$ 1.3) $\times$ 10$^{13}$ cm$^{-2}$ and $N$(HDS)/$N_{\mathrm{H_2}}$ = 8 $\times$ 10$^{-10}$ (assuming $T_{\mathrm{rot}}$ = 12 K). 
Abundances of HDS in Galactic star-forming cores are typically in the range of 10$^{-11}$--10$^{-10}$ according to \citet{Vas03}. 
The estimated HDS abundance at the East core is comparable to typical Galactic values despite the lower elemental abundances in the SMC. 
Further observations of other transitions and related molecular species are highly required to understand deuterium chemistry in low-metallicity environments.

\section{Summary} \label{sec_sum} 
We report the results of 0.1 pc scale submillimeter observations towards a high-mass young stellar object (IRAS 01042-7215) in the Small Magellanic Cloud with ALMA. 
Complementary infrared data obtained with \textit{AKARI} and the Gemini South telescope are also presented. 
The following conclusions are obtained in this work: 

\begin{enumerate}
%
%Detections
\item 
Emission lines of CS, C$^{33}$S, H$_2$CS, SO, SO$_2$, CH$_3$OH, H$^{13}$CO$^{+}$, H$^{13}$CN, SiO, and dust continuum are detected. 
Among these species, CH$_3$OH, SO$_2$, H$_2$CS, C$^{33}$S, and H$^{13}$CN are for the first time detected in the SMC. 
Tentative detection of HDS is also reported. 

%Physical properties
\item 
The target source is spatially resolved into two dense molecular cloud cores, named the YSO core and the East core. 
The YSO core is very dense ($n_{\mathrm{H_2}}$ $\sim$10$^6$ cm$^{-3}$) and well shielded ($A_{V}$ $\sim$10--30 mag), and is associated with an embedded high-mass YSO. 
The East core is also dense ($n_{\mathrm{H_2}}$ $\sim$10$^5$ cm$^{-3}$) and shielded ($A_{V}$ $\sim$4--10 mag), but is not associated with an infrared source. 
A gas temperature in the East core is estimated to be $\sim$10 K based on the rotation analysis of CH$_3$OH lines. 
A possible Galactic counterpart for the East core would be a massive starless core, an embedded high-mass YSO(s) before the emergence of infrared emission, or a cluster of low-mass embedded YSOs. 

%Chemical properties
\item 
The above two dense cores, which are separated by $\sim$0.3 pc, show different chemical characteristics. 
The chemical composition of the YSO core is characterized by the low abundance of molecular gas and the presence of ices. 
In contrast, the East core is characterized by relatively rich molecular lines, and the species except for CS, SO, and H$^{13}$CO$^{+}$ are detected only in the East core. 
Ices are not detected toward the East core in the current data owing to the lack of a bright infrared continuum source. 
We speculate that the effect of shock chemistry and/or the different degree of the adsorption of gas-phase species onto dust may contribute to the chemical differentiation.

%Methanol 
\item
The fractional abundance of CH$_3$OH gas in the East core is estimated to be (0.5--1.5) $\times$ 10$^{-8}$, which is comparable with or marginally higher than those of similar cold sources in our Galaxy despite a factor of five lower metallicity in the SMC. 
This would indicate an enhanced production or a suppressed destruction of gas-phase CH$_3$OH in the present SMC source compared to Galactic analogues. 
Although it is still unclear whether the observed source represents common characteristics of cold and dense molecular gas in the SMC, this work provides observational evidence that an organic molecule like CH$_3$OH, which is largely formed on grain surfaces, can be produced even in a significantly lower metallicity environment compared to the solar neighborhood. 
A possible origin of cold CH$_3$OH gas in the East core is discussed. 
\end{enumerate}

\acknowledgments 
This paper makes use of the following ALMA data: ADS/JAO.ALMA$\#$2016.1.00394.S. 
ALMA is a partnership of ESO (representing its member states), NSF (USA) and NINS (Japan), together with NRC (Canada) and NSC and ASIAA (Taiwan) and KASI (Republic of Korea), in cooperation with the Republic of Chile. 
The Joint ALMA Observatory is operated by ESO, AUI/NRAO and NAOJ. 
This work has made extensive use of the Cologne Database for Molecular Spectroscopy. 
This work uses data based on observations with \textit{AKARI}, a JAXA project with the participation of ESA. 
This work also uses data obtained at the Gemini South telescope as a part of the time exchange program between the Gemini and Subaru telescopes (Program ID: S10B-120). 
We also use data obtained by the Two Micron All Sky Survey project, NASA's \textit{Spitzer} Space Telescope, and ESA's \textit{Herschel} Space Observatory. 
We are grateful to all the members who contributed to these projects. 
This work is supported by a Grant-in-Aid from the Japan Society for the Promotion of Science (15K17612). 
T. Shimonishi was supported by the ALMA Japan Research Grant of NAOJ Chile Observatory, NAOJ-ALMA-183. 
Y.N. was supported by NAOJ ALMA Scientific Research grant Number 2017-06B. 
Finally, we would like to thank an anonymous referee for insightful comments, which substantially improved this paper.

\end{document}